\RequirePackage{fix-cm}
\documentclass[smallextended]{svjour3}       

\smartqed  
\usepackage{graphicx}
\usepackage{graphicx,epstopdf}
 \usepackage{mathptmx}      
\usepackage{latexsym}
\usepackage{amssymb}
\usepackage{amsmath}
\usepackage{stmaryrd}
\begin{document}

\title{Linear Whitham-Boussinesq modes in channels of constant cross-section.
}


\author{R. M. Vargas-Maga\~na\and P. Panayotaros\and A.A. Minzoni}


\institute{R. M. Vargas-Maga\~na \at Mathematical Sciences Research Institute
               17 Gauss Way Berkeley, CA 947205070\\
              Tel.: +510-643-6844\\
              \email{rosavargas@berkeley.edu}           
           \and
           P. Panayotaros \at
           Departamento de Matem\'aticas y Mec\'anica IIMAS, Universidad Nacional Aut\'onoma de M\'exico,
           Apdo. Postal 20-726, 01000 Cd. M\'exico,
M\'exico
}

\date{Received: date / Accepted: date}

\maketitle

\begin{abstract}

  We study normal modes for the
  linear water wave problem in infinite straight
channels of bounded constant cross-section.
Our goal is to compare
semi-analytic normal mode solutions known in the literature
for special triangular cross-sections, namely isosceles triangles of equal angle of
$45^{\circ}$ and  $60^{\circ}$, see Lamb \cite{lamb1932hydrodynamics}, Macdonald \cite{macdonald1893waves} ,
Greenhill \cite{greenhill1887wave},
Packham \cite{packham1980small}, and Groves \cite{groves1994hamiltonian},
to numerical solutions obtained using
approximations of the non-local Dirichlet-Neumann operator
for linear waves, specifically an ad-hoc approximation
proposed in \cite{vargas2016whitham}, and a first order truncation of the 
systematic depth expansion by Craig,
Guyenne, Nicholls and Sulem \cite{CGNS}.
We consider cases of transverse (i.e. 2-D) modes and longitudinal modes, i.e.
3-D modes with sinusoidal dependence
in the longitudinal direction.
The triangular geometries considered have slopping beach boundaries that should in principle
limit the applicability of the approximate Dirichlet-Neumann operators. We nevertheless
see that the approximate operators give remarkably close results for transverse even modes,
while for odd transverse modes we have some discrepancies near the boundary.
In the case of longitudinal modes, where the theory only yields even modes,
the different approximate operators show more discrepancies for the first two
longitudinal modes and better agreement for higher modes.
The ad-hoc approximation is generally closer to exact modes away from the boundary.

\keywords{Linear water waves \and Whitham-Boussinesq model over variable topography
  \and Dirichlet-Neumann operator \and Normal modes on triangular straight channels \and pseudodifferential operators}
\end{abstract}

\section{Introduction}
\label{intro}

We study some problems on normal modes in the linear water wave theory,
in particular we compute  
transverse and longitudinal normal modes in infinite straight channels of constant bounded cross-section,
considering special
depth profiles for which there are known
explicit or semi-explicit analytical solutions in the literature
\cite{greenhill1887wave,groves1994hamiltonian,lamb1932hydrodynamics,macdonald1893waves,packham1980small}.   
We compare the results to numerical normal modes obtained by using
simple approximations of the nonlocal variable depth Dirichlet-Neumann operator
for the linear water wave equations, in particular a recently proposed
ad-hoc approximation \cite{vargas2016whitham}, as well as
a first order truncation of the systematic expansion in the depth proposed by Craig,
Guyenne, Nicholls and Sulem \cite{CGNS}.
Our main motivation is to test the simple approximations to the
Dirichlet-Neumann operator
as they are of interest 
in constructing simplified nonlocal shallow water models, e.g.
Whitham-Boussinesq equations \cite{AcevesSanchez201380,Carter2018,HurTao2018,moldabayev2014whitham} and     \textit{arXiv:1608.04685} submitted manuscript by V. M. Hur and A. K. Pandey.
Such models have attracted
considerable current interest
\cite{constantin1998wave,ehrnstrom2009traveling,ehrnstrom2012existence,hur2015breaking,naumkin1994nonlinear} and  \textit{arXiv:1602.05384} submitted manuscript by M. Ehrnstrom  and E. Wahl{\'e}n.
The inclusion of variable depth effects
raises additional questions
on the dynamics of these systems
and is of interest
in geophysical and coastal engineering applications.

The nonlocal linear system we use to compute normal modes is derived using the Hamiltonian
formulation of the free surface potential flow \cite{zakharov1968stability}, see also
\cite{miles1977hamilton,radder1992explicit}, and approximations for the   
(nonlocal) Dirichlet-Neumann operators for the Laplacian in the fluid domain
appearing in the
kinetic energy part of the Hamiltonian \cite{craig1994hamiltonian}.
Explicit infinite series
expressions for the Dirichlet-Neumann operator in variable depth
were derived by Craig, Guyenne, Nicholls and Sulem \cite{CGNS}, see also \cite{lannes2013water}.
Such expressions are generally complicated and 
can be used for numerical computations \cite{gouin2015development},
or to further simplify the equations of motion.
The Hamiltonian and Dirichlet-Neumann formulation implies that 
variable depth effects are already captured
at the level of the linear theory, i.e. the nontrivial wave-amplitude
operator is expressed recursively in terms of the
linear operator \cite{craig1993numerical,CGNS}. The study of linear normal modes is therefore  
a good test problem for comparing different approximations
of the Dirichlet-Neumann operator for variable depth.
In this paper
we focus on simple approximations of the variable depth operator, such as
the ad-hoc generalization of the constant depth operator
for arbitrary depth proposed in \cite{vargas2016whitham}. This operator has 
some of the structural properties of the exact Dirichlet-Neumann operator, e.g. symmetry and
exact infinite depth asymptotics, but is also seen to lead to a depth expansion
that differs from the exact one of \cite{CGNS}.

The idea of the paper is compare results obtained 
using the approximate Dirichlet-Neumann operators to
semi-analytic normal mode solutions known
for some special depth profiles. These analytical results rely on the existence of
families of harmonic functions that satisfy
the rigid wall boundary conditions at the bottom,
and can be only obtained for a few special depth profiles,
such as isosceles triangles
with sides inclined at $45^{\circ}$ and  $60^{\circ}$ to the vertical, see
Greenhill  \cite{greenhill1887wave}, Macdonald  \cite{macdonald1893waves},
and the summary in Lamb's book, \cite{lamb1932hydrodynamics}. More recent studies are by
Packham  \cite{packham1980small} and Groves \cite{groves1994hamiltonian}.
A complete set of modes was also obtained for a semicircular channel
by Evans and Linton \cite{evans1993sloshing}.
The construction typically yields even and odd normal modes that we then
compare to even and odd eigenfunctions
of approximate Dirichlet-Neumann operators in a periodic domain.

One problem with our plan is
that the non-constant depth examples with classical analytic solutions we are aware of
concern domains with
a slopping beach, and as we clarify below, the classical and periodic Dirichlet-Neumann approaches
are not equivalent because of the
different assumptions at the intersection between the horizontal and sloping beach boundaries.
Despite this problem, we find that the two approaches
give comparable and often
very close results for the 2-D normal mode shapes, especially away from the beach.
This is especially
the case for even modes, where we see good agreement in the entire domain.
Results for odd modes are close away from the
beach, but have a marked discrepancy near the beach. In that case the periodic problem
for the approximate Dirichlet-Neumann operators leads to
modes with that vanish at boundary, while odd exact modes have maxima at the boundary.
In the problem of 3-D even longitudinal modes, we consider moderate speed along the transverse direction and
see more discrepancies between the exact and approximate modes. Agreement in the interior is good
after the first two modes, but discrepancies at the boundary persist for higher even modes.
The approximate Dirichlet-Neumann approach
also gives results for the odd case, where there are no exact results. 

In summary, the approximate periodic Dirichlet-Neumann operators give better results for
the transverse 2-D modes, especially in the interior.
In \textit{Section 4}
we present evidence that the modes computed by the approximate
periodic Dirchlet-Neumann operators are limiting cases of modes corresponding to a periodic depth
profiles with nowhere vanishing depth.
This observation explains intuitively why the approximate operators can not capture
the slopping beach rigid wall boundary condition as realistically as the classical exact approach.

The organization of the paper is as follows.
In \emph{Section 2} we formulate the water wave problem using the Dirichlet-Neumann
operators and present the operators used to approximate the linear system.
In \textit{Section 3} we formulate the classical problem of transverse and longitudinal modes
for the linear theory and define analogues that use approximate Dirichlet-Neumann
operators. In \textit{Section 4} we present our results, comparing normal modes
obtained using the classical and approximate Dirichlet-Neumann approaches.
In \textit{Section 5} we briefly discuss the results.

\section{Water wave problem in variable depth and approximate Dirichlet-Neumann operators}
\label{sec:1}

Following the
Hamiltonian formulation of the water wave problem due to Zakharov \cite{zakharov1968stability},
see also \cite{miles1977hamilton,radder1992explicit},
the Euler equations for free surface potential flow
can be restated as a Hamiltonian system in terms of
the wave amplitude $\eta(x,t)$ and surface hydrodynamic
potential $\xi(x,t)=\varphi(x,\eta(x,t),t),$ namely as
\begin{equation}\label{hamiltons-equations}
\partial_t  \left( \begin{array}{l}
\eta  \\
\xi \\
\end{array} \right) =\left( \begin{array}{cc} 0& I\\ -I& 0  \end{array} \right)   \left( \begin{array}{l}
\frac{\delta  H}{\delta \eta} \\
\frac{\delta  H}{\delta \xi} \\
\end{array} \right),
\end{equation}
where the Hamiltonian is expressed explicitly in terms of $\eta$ and $\xi$ as
\begin{equation}\label{ham}
H = \frac{1}{2} \int_{\mathbb{R}} (\xi G(\beta,\eta)\xi + g\eta^2)dx,
\end{equation}
see \cite{craig1994hamiltonian}, and the operator $G(\beta, \eta)$ is defined as follows:
consider the elliptic problem
\begin{eqnarray}\label{ecsup}
\Delta\varphi(x,y)= 0, &  &   \forall \text{ } (x, y) \in \mathcal{D}_t(\eta) \\
\varphi(x, \eta(x)) = \xi(x), &  & \forall  \text{ }  x \in \mathbb{R}, \\
\frac{\partial\varphi}{\partial \hat{n}}(x, -h_0+\beta(x)) = 0, & & \forall \text{ }  x \in \mathbb{R},
\end{eqnarray}
in the two dimensional (time-dependent) simply connected domain
$\mathcal{D}_t(\eta):= \lbrace (x,y): x\in \mathbb{R}, -h_0 + \beta(x) < y < \eta(x,t)\rbrace$.
If $\eta$ and $\xi$ are sufficiently smooth and decay at infinity then (\ref{ecsup})-(5) admits
a unique solution and we can compute the normal derivative of the solution at the surface $y=\eta$.
The Dirichlet-Neumann operator $G(\beta, \eta)$ is then defined by
\begin{equation}
(G(\beta,\eta)\xi)(x)=(1+ (\partial_x\eta(x))^2)^{\frac{1}{2}} \triangledown\varphi(x, \eta(x)) \cdot N(\eta(x)),
\end{equation}
where  $N(\eta(x))= (1+ (\partial_x\eta(x))^2)^{-\frac{1}{2}} (-\partial_x\eta(x),1)$, $x \in \mathbb{R}$, 
is the exterior unit normal at  the free surface. The Dirichlet-Neumann operator $G(\beta, \eta)$ is 
a linear operator on $\xi$ and is symmetric with respect to the usual $L^2$ inner product. Similar definitions apply
to the periodic problem and to higher dimensions.

In  \cite{CGNS}, Craig, Guyenne, Nicholls and Sulem give an expansion of this operator
in the presence of non-trivial bottom topography
\begin{equation} \label{Ganalytic}
G(\beta, \eta)
= G_0(\beta, \eta) +G_1(\beta, \eta)+G_2(\beta, \eta)+ ...,
\end{equation}
\noindent where the  $G_j$ are homogeneous of degree $j$ in $\eta.$ The first terms are
\begin{eqnarray}\label{expanalit}
G_0(\beta, \eta)& = &  D\tanh(h_0D) + DL(\beta), \\
G_1(\beta, \eta) & = &  D \eta D - G_0\eta G_0, \\
G_2(\beta, \eta) & = & \frac{1}{2} (G_0 D \eta ^2 D- D^2 \eta ^2 G_0-2 G_0  \eta G_1),
\end{eqnarray}
\noindent where  $D=-i\partial_x$, and $G_0 = G_0[\beta, \eta] $. We are
also using the notation
\begin{equation} \label{OPsigma-1}
[a(f(x) D^m) \xi ](x)  = \frac{1}{2 \pi} \int_{\mathbb{R}} a(f(x) k^m) \hat{\xi}(k) e^{i k x} d k,
\end{equation}
with $a$, $f$ be real functions, and
$\hat{\xi}(k)= \int_{\mathbb{R}} \xi(x) e^{-ik x} d k$,
the Fourier transform of the real function $\xi$.

At higher order,
the $G_j$, $j > 2$, are similarly obtained from $G_0$, using a recursion formula.
The recursion formula for the $G_{j}$ is similar to the one obtained for a flat bottom, where
$G_{0} = D\tanh(h_0D)$, see \cite{craig1993numerical}.
Variable depth effects are thus encoded in the operator $L(\beta)$.

The operator $L(\beta)$ can be expressed in powers of the depth variation $\beta$ as
$L(\beta) = \Sigma_{j=0} ^{\infty} L_j(\beta),$  where the $L_j(\beta)$ are  homogeneous
of order $j$  in $\beta$, and are computed recursively, see  \cite{CGNS}.
The first two terms  in the expansion are
\begin{eqnarray}\label{expL}
\label{L-operator-expansion-1}
L_0(\beta) &=& 0,\\
\label{L-operator-expansion-2}
L_1(\beta) &=& -\text{sech} (h_0D)\beta D \text{sech}(h_0D).
\end{eqnarray}
The first two terms of this expansion lead the
first order
approximation $\mathcal{A}_1$
of the Dirichlet-Neumann operator
\begin{equation}\label{ExpCS}
\mathcal{A}_1(\beta) = D\tanh(h_0D)  -D\text{sech}(h_0D) \beta D\text{sech}(h_0 D)
\end{equation}
that we use below.
This operator was used recently by W. Craig, M. Gazeau, C. Lacave, C. Sulem to calculate bands for periodic depth variation see \textit{arXiv:1706.07417} submitted manuscript.

Higher order expansions in $\beta$ have been considered in the
numerical study of \cite{gouin2015development}. To avoid these longer expressions
in simplified nonlocal shallow water equations,
\cite{vargas2016whitham} proposed an ad-hoc approximation
$\mathcal{A}_{G_0}$ of the linear Dirichlet-Neumann operator given by
\begin{equation}\label{AG0}
\mathcal{A}_{G_0}(\beta) =\textit{Sym}(D\tanh(h(x)D)),
\end{equation}
where $h(x)=h_0-\beta(x)$.
The (formal) symmetrization of a linear operator  
$\mathcal{A}$ in $L^2 = L^2(\mathbb{R};\mathbb{R})$
with the (standard) inner product $\langle f,g \rangle = \int_{\mathbb{R}} f(x)g(x)dx$,
is defined by
$\textit{Sym} (\mathcal{A}) = \frac{1}{2}( \mathcal{A} +  \mathcal{A}^*)$, where $\mathcal{A}^*$ is
the adjoint of $\mathcal{A}$ with respect to the
inner product, i.e. $\langle \mathcal{A} f,  g \rangle = \langle f, \mathcal{A}^* g \rangle$, for all $f$, $g$
in the the domains of $\mathcal{A}$, $\mathcal{A}^*$ respectively.
(It is assumed that the domain $D(\mathcal{A})$ of $\mathcal{A}$ is dense in $L^2$.)
$\mathcal{A} $ is symmetric if $D(\mathcal{A}) = D(\mathcal{A}^*)$ and $ \mathcal{A} =  \mathcal{A}^*$,
thus $D(\mathcal{A}) = D(\mathcal{A}^*)$ implies that
$\textit{Sym} (\mathcal{A}) $ is symmetric.

Symmetrizing an approximate Dirichlet-Neumann operator is natural by \eqref{hamiltons-equations}, \eqref{ham},
e.g. $H = \frac{1}{2} \int_{\mathbb{R}} (\xi \mathcal{A} \xi + g\eta^2)$, $\mathcal{A}$ a linear operator, leads formally to the
equations ${\partial_t \eta} =  \frac{1}{2}(\mathcal{A} +  \mathcal{A}^*) \xi$, ${\partial_t \xi} = - g \eta$. 
We note that $D\tanh(h(x)D)$ maps real-valued functions to real-valued functions,
see Appendix A and the next section for further details on symmetrizing this operator.

We check that $\mathcal{A}_1(\beta)$ of \eqref{ExpCS} is symmetric.
To compare this operator to $\mathcal{A}_{G_0}$, we
expand $D\tanh(h(x)D)$ in $\beta$ to $O(\beta^2)$, and symmetrize. The result is the operator
\begin{eqnarray}\label{ExpansionAproxAG0Texto}
  \mathcal{A}_2(\beta)&=& D\tanh(h_0D) + \frac{1}{2} [ i\beta'\text{sech}^2(h_0D)D - \beta \text{sech}^2(h_0 D) D^2 \\  \nonumber
    & & - iD\text{sech}^2(h_0D) \beta' + D^2 \text{sech}^2(h_0D) \beta],
\end{eqnarray}
see Appendix A. We see that operators $\mathcal{A}_1$ and $\mathcal{A}_2$ are  apparently different.


As was pointed out in \cite{CGNS},
the formulation can be extended to 3-D domains. Recently in \cite{andrade2018three} a Dirichlet-Neumann operator for a three-dimensional surface water wave problem in the presence of highly variable, non smooth  topagraphy is constructed  using a Galerkin method in Fourier space.
\medskip

Let $a$, $f$ be real functions and $g:{\mathbb{R}}^2 \rightarrow {\mathbb{R}}$,
and let $D=(D_1,D_2)^T=-i(\partial_{x_1},\partial_{x_2})^T$. We then define operators
$a(f(x)g(D))$ by
\begin{equation} \label{OPsigma-2}
  [a(f(\mathbf{x}) g(D)) \xi ](\mathbf{x})  =
  \frac{1}{(2 \pi)^2} \int_{\mathbb{R}^2} a(f(\mathbf{x}) g(\mathbf{k}))
  \hat{\xi}(\mathbf{k}) e^{i \mathbf{k} \cdot \mathbf{x}} d^2 \mathbf{k},
\end{equation}
with $ {\hat \xi}(\mathbf{k})= \int_{\mathbb{R}^2} {\xi}(\mathbf{x}) e^{-i \mathbf{k} \cdot  \mathbf{x}} d^2 {\mathbf{x}} $,
the Fourier transform of the function $\xi$ on $ {\mathbb{R}}^2 $.

Letting 
\begin{equation}
\left|  D\right|  =\sqrt{\left| D_1\right|^2+ \left| D_2\right| ^2},
\end{equation}
the expansion of $L(\beta)$ up to $O(\beta^2)$ leads to the approximate
Dirichlet-Neumann operator
\begin{equation}\label{A1en3D}
  \mathcal{A}_1(\beta)= [\left| D\right|\tanh(h_0\left| D\right|)]  -
          [\left| D\right|\text{sech}(h_0\left| D\right|)][\beta(\textbf{x}) \left| D\right|\text{sech}(h_0 \left| D\right|)].
\end{equation}
We will also consider the 3-D analogue of the ad-hoc operator of (\ref{AG0})
\begin{equation}\label{Aproxgorro}
\mathcal{A}_{G_{0}}(\beta)= \textit{Sym}[ \left| D\right|\tanh(h(\mathbf{x})\left| D\right| )], \text{ } \text{}
\end{equation}
with $ h(\mathbf{x})=h_0 - \beta(\mathbf{x} )$. The symmetrization is defined as in $\mathbb{R}$.

%

\section[Formulation of the problem]{Linear modes in channels and
    \\ Dirichlet-Neumann operators}\label{formulation}
\label{sec:2}

We now consider the classical formulation of the problem of
linear modes in channels.
We use Cartesian coordinates
denoted by $(x,y, z)$, where $y$ is directed vertically upwards,
$x$  is measured longitudinally along the channel and $z$
is measured across the channel, see e.g. \figurename{ \ref{Figbeta45}} and  \ref{Figbeta30}

We define the fluid domain  as
$\mathcal{D}= \mathbb{R} \times \Omega,$ where $\Omega \subset \mathbb{R}^2$ is the cross section.
We assume a bounded cross sections $\Omega = \Omega_B$ of the form
\begin{equation}\label{omega-bounded}
\Omega_B = \lbrace [z,y]: z \in [0,b], y \in  [h_m+\beta(z), h_M],
\end{equation}
The heights $y = h_m$ and $y = h_M$ describe the minimum and maximum elevations of the
fluid domain respectively, with
$h_m < h_M$, and $h_m + \beta(z) \leq h_M$ for all $z$.

We will assume
$\partial \mathcal{D}= \mathbb{R} \times \partial \Omega =\Gamma_{L} \cup  \Gamma_{F} \cup \Gamma_{B}$,
with $\Gamma_L$ representing the lateral wall,
$\Gamma_F$ representing the free surface, and $\Gamma_B$ the bottom,
\begin{equation}
\Gamma_F= \lbrace (x,h_M,z): x\in \mathbb{R}, 0\leq z \leq b \rbrace,
\end{equation}
\begin{equation}
\Gamma_B= \lbrace (x, h_{m}+\beta(z), z): x\in \mathbb{R}, 0\leq z \leq b  \rbrace,
\end{equation}
\begin{equation}
\Gamma_L= \lbrace (x,y,0): x\in \mathbb{R}, y \in [h_m+ \beta(0), h_M]\rbrace \cup
\lbrace (x,y,b): x\in \mathbb{R}, y \in [h_m+ \beta(b), h_M]\rbrace.
\end{equation}
In this article we are interested in domains with $h_m + \beta(z) = h_M$ at $z=0$ and $z = b$.
Then $\Gamma_L= \emptyset$.

To state the problem we introduce a velocity potential $\phi(x,y,z,t)$
and look for solutions of  Laplace's equation
\begin{equation}\label{Lap}
\phi_{xx}+ \phi_{yy} +\phi_{zz}=0 \text{ in } \mathcal{D,}
\end{equation}
with
\begin{equation}
\label{Ebc}
\left\lbrace
\begin{array}{l}
\phi_t + g\eta= 0 \text{ at } \Gamma_F,  \\
\eta_t= \phi_y \text{ at } \Gamma_F,\\
\frac{\partial \phi}{\partial {\hat n}}=0 \text{ at } \Gamma_B \cup \Gamma_L,
\end{array} \right.
\end{equation}
see \cite{whitham2011linear}. Equations
\eqref{Lap}, \eqref{Ebc}
are the linearized Euler equations
for free surface potential flow.
    \begin{figure}
            \label{fig:T45}       
        \includegraphics[scale=0.3]{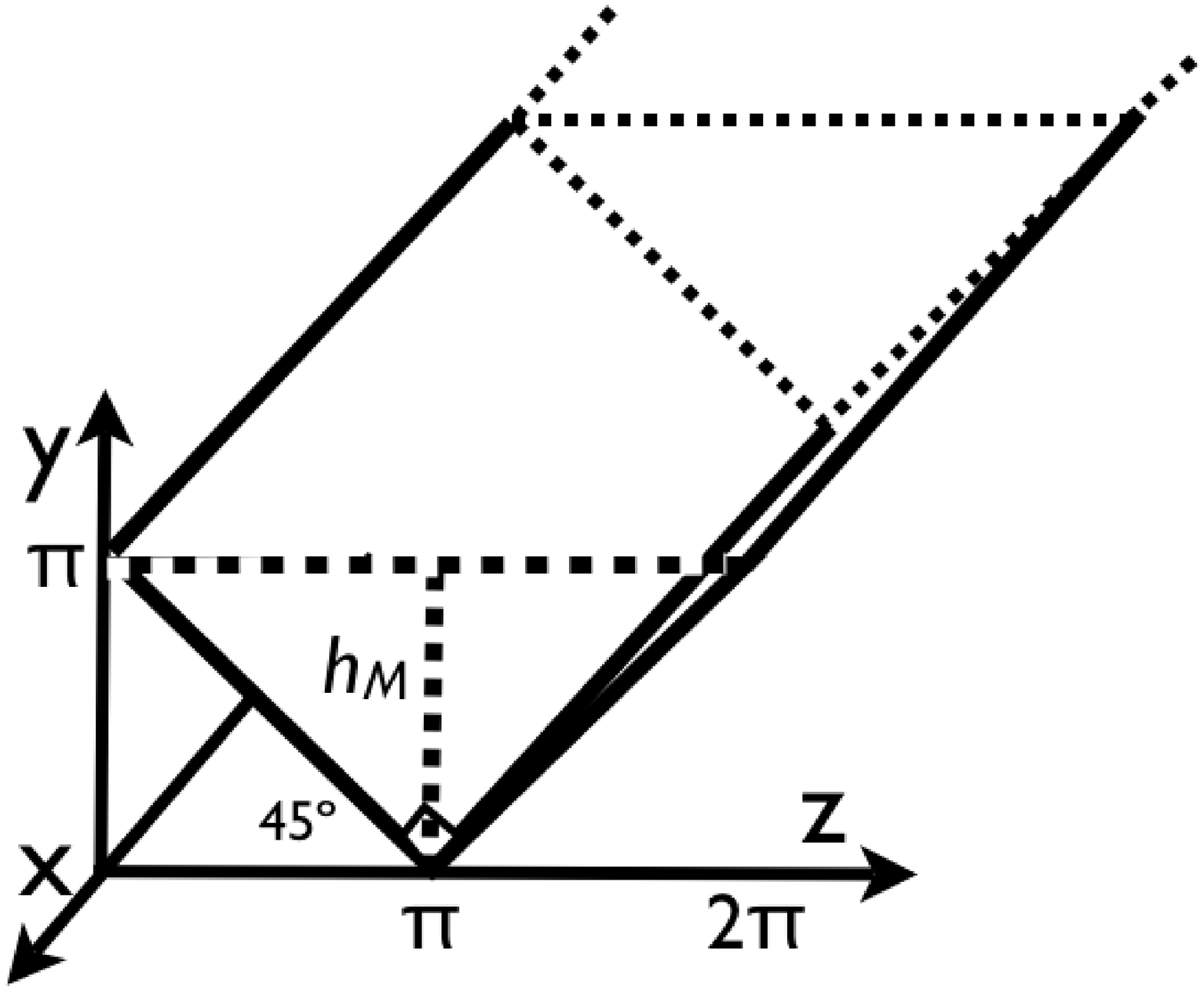}
        \caption{Schematic of straight channel with triangular cross-sections $\Omega=\beta_{45}(z)$ and coordinate system}   
    \end{figure}
    \begin{figure}
            \label{fig:T30}       
        \includegraphics[scale=0.25]{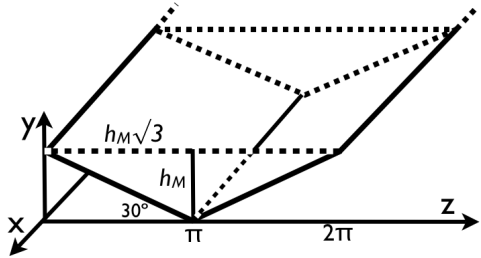}
        \caption{Schematic of straight channel with triangular cross-sections $\Omega=\beta_{30}(z)$ and coordinate system}
    \end{figure}

We consider solutions of the following two forms:

\noindent A.
\begin{equation}\label{TransverseMode}
\phi (x,y,z,t)=\psi(y,z)\cos(\omega t),
\end{equation}
referred to as \textit{transverse  modes}  see \cite{packham1980small}, \cite{groves1994hamiltonian}.

\noindent B.
\begin{equation}\label{LongMode}
\phi(x,y,z,t)=\psi(y,z)\cos(\kappa x-\omega t),
\end{equation}
referred to as \textit{longitudinal  modes}, see \cite{packham1980small}, \cite{kuznetsov2002linear}.

We formulate the problem of finding solutions of the above form in terms of Dirichlet-Neumann operators.

We first consider transverse modes.
Let the fluid domain consist of a straight channel and consider the case $\Omega = \Omega_{B}.$
Consider $f: [0,b] \rightarrow \mathbb{R}$, and define
the Dirichlet-Neumann operator
$G_{0}$ by
\begin{equation}\label{OpGo}
(G_0 f)(z) =\frac{\partial \psi(z,y)}{\partial y} |_{y = h_M },
\end{equation}
where $\psi: \Omega \rightarrow \mathbb{R}$ satisfies
\begin{equation}\label{EqUPplugging}
\left\lbrace \begin{array}{l}
\Delta\psi= 0\text{ } \text{ in }\Omega, \\
\psi = f  \text{ at } \Omega \cap \Gamma_F , \\
\frac{\partial \psi}{\partial {\hat n}}=0 \text{ } \text{ at }\Omega \cap (\Gamma_B \cup \Gamma_L), \\
\end{array} \right.
\end{equation}
Combining the first two equations of \eqref{Ebc},
the problem of finding
transverse mode solutions \eqref{TransverseMode}
can be written as
\begin{equation}
\label{DN-eigenfunctions}
G_0 f = g^{-1}\omega^2 f.
\end{equation}

The classical (exact) approach to solving \eqref{Lap}-\eqref{TransverseMode},
or \eqref{OpGo}-\eqref{DN-eigenfunctions}, is given in subsection 4.1.
We will compare the results to solutions of \eqref{DN-eigenfunctions}
with $G_0$ replaced by the operators  $\mathcal{A}_1(\beta)$, $\mathcal{A}_{G_0}(\beta)$,
with depth topography as in \eqref{omega-bounded},
applied to $b-$periodic functions.
We comment on this approximation at the end of this section.

We now consider longitudinal modes.
Consider $f(z)=\psi(0,z)$,
with $z \in [0,b],$
and define the
modified Dirichlet-Neumann operator $G_{0,\kappa}$ by
\begin{equation}\label{operadorgorro}
(G_{0,\kappa} f)(z) =  \frac{\partial \psi(z,y)}{\partial y} |_{y = h_M },
\end{equation}
where $\psi : \Omega \rightarrow \mathbb{R}$ satisfies
\begin{equation}\label{EqTrans}
\left\lbrace \begin{array}{l}
\Delta\psi=\kappa^2 \psi \text{ } \text{ in }\Omega, \\
\psi = f \text{ at } \Omega \cap \Gamma_F , \\
\frac{\partial \psi}{\partial {\hat n}}=0 \text{ } \text{ at }\Omega \cap (\Gamma_B \cup \Gamma_L), \\
\end{array} \right.
\end{equation}
with $\kappa$ as in (\ref{LongMode}).
By \eqref{Lap} and the first two equations of \eqref{Ebc},
the problem of finding longitudinal mode
solutions \eqref{LongMode}
can be written as
\begin{equation}\label{modified-DN-eigenfunctions}
G_{0,\kappa} f  = g^{-1} \omega^2 f.
\end{equation}
Comparing \eqref{Lap}, \eqref{LongMode}, and the first equation of \eqref{EqTrans},
$(G_{0,\kappa} f)(z)$ in \eqref{modified-DN-eigenfunctions} is the 3-D zeroth-order Dirichlet-Neumann operator,
$ G_0$ applied to functions $f(z)e^{\pm i \kappa x}$
(or $f(z)\cos(\kappa x))$.

The classical (exact) approach to solving \eqref{Lap}-\eqref{Ebc} and \eqref{LongMode},
or \eqref{operadorgorro}-\eqref{modified-DN-eigenfunctions},
is given in subsection 4.2.
The solutions will be compared to solutions of \eqref{modified-DN-eigenfunctions}
with $G_{0,\kappa}$ replaced by 
operators $\mathcal{A}_1$ of \eqref{A1en3D}, $\mathcal{A}_{G_0}$ of \eqref{Aproxgorro},
applied to functions of the form $ e^{\pm i \kappa x}f(z)$, with
$f$ $b-$periodic.
The depth topography is as in \eqref{omega-bounded}, and $\beta$ is independent of $x$.
Specifically, the operator $\mathcal{A}_1$ of
\eqref{A1en3D} applied on functions of the form $ e^{\pm i \kappa x}f(z)$
defines the operator $\mathcal{A}_{1,\kappa}(\beta)$ by  
\begin{eqnarray}\label{A1en3D-kappa}
 \mathcal{A}_{1,\kappa}(\beta) f& = & [\sqrt{D^2 + \kappa^2}  \tanh(h_0 \sqrt{D^2 + \kappa^2}) \\  \nonumber
          &  & -\sqrt{D^2 + \kappa^2} \text{sech}(h_0 \sqrt{D^2 + \kappa^2})\beta(z)
            \sqrt{D^2 + \kappa^2} \text{sech}(h_0 \sqrt{D^2 + \kappa^2})]f,
\end{eqnarray}
with $ h(z)=h_0 - \beta(z) $, $h_0 = h_M - h_m >  0$, and $D= -i\partial_z$.
We will compute numerically the eigenmodes of $\mathcal{A}_{1,\kappa}(\beta)$ with $b-$periodic boundary conditions.
Similarly, the ad-hoc operator $\mathcal{A}_{G_0}$ of \eqref{Aproxgorro}
applied on functions of the form $ e^{\pm i \kappa x}f(z)$
defines the operator
$\mathcal{A}_{G_{0,\kappa}}(\beta)$ by
\begin{equation}
  \label{Aproxgorro-kappa}
\mathcal{A}_{G_{0,\kappa}}(\beta) f= [\textit{Sym}(\sqrt{D^2 + \kappa^2} \tanh(h(z)\sqrt{D^2 + \kappa^2}))]f, \text{ } \text{}
\end{equation}
with the notation of \eqref{A1en3D-kappa}, and we will compute
the eigenmodes of $\mathcal{A}_{G_{0,\kappa}}(\beta)$ with $b-$periodic boundary conditions.
Symmetrization of operators on $b-$periodic functions is as in \emph{Section 2}, using
the standard inner product on real $b-$periodic $L^2$ functions.

Computations with $b-$periodic boundary conditions use the periodic analogues of the operators
of \emph{Section 2} and above.
In particular for $a$ a real function
and $h$, $r$ real $b-$periodic functions, we let
\begin{equation}\label{PDOazD}
[a(h(z)D)f](z)
=\sum \limits_{\lambda=-\infty}^{\infty}  a(h(z)\lambda) \hat{f}_\lambda  e^{i \lambda\frac{2 \pi}{b}z},
\end{equation}
\begin{equation}\label{PDOazD-Fourier}
\hat{f}_\lambda= \frac{1}{b} \int_{-{\frac{b}{2}}}^{\frac{b}{2}} f(z) e^{-i\lambda\frac{2\pi}{b}z}dz.
\end{equation}

Numerical computations use Galerkin truncations of the above operators to modes $|l| \leq l_{max}$.
We can simplify the calculations
by noting that operators $\mathcal{A}_1$, $D\tanh(h(\cdot)D)$, $\mathcal{A}_2$ of \emph{Section 2} (with )
and $\mathcal{A}_{1,\kappa}(\beta)$, $\mathcal{A}_{G_{0,\kappa}}(\beta)$ above map real-valued functions
to real-valued functions, and for $\beta$, $h$ even they also preserve parity, see Appendix A.
We can thus consider finite dimensional truncations of expansions in cosines and sines
for the even and odd subspaces respectively. The corresponding matrices are
block-diagonal. In the case of
$D\tanh(h(\cdot)D)$, we compute the matrix
for $\textit{Sym}(D\tanh(h(\cdot)D))$ by symmetrizing the truncation of
$D\tanh(h(\cdot)D)$. Clearly, the adjoint (transpose) and symmetrization also preserve parity.

In the next section we describe some known
semi-analytic solutions of the form \eqref{TransverseMode}, \eqref{LongMode},
obtained for some special domains with finite cross-section
$\Omega$ and $\Gamma_L = \emptyset$, i.e. slopping beach geometries.
These solutions are constructed starting with a multiparameter
family of harmonic functions $\phi_\mu$ of the form
\eqref{TransverseMode}, or \eqref{LongMode},
defined on the plane and satisfying the rigid wall boundary condition on a
set ${\tilde \Gamma}_B$ that includes
$\Gamma_B$ and
is the boundary of a domain
$\tilde{\mathcal{D}}$ that includes $\mathcal{D}$.
Then we require that the $\phi_\mu$ also satisfy the
first two equations of \eqref{Ebc} on the free surface $\Gamma_F$.
This requirement leads to algebraic equations
that restrict the allowed values of the parameter $\mu$
to a discrete set and also determine the
frequencies. This construction does not assume any boundary conditions
for the free surface potential.

The second computation in the next section computes $b-$periodic eigenfunctions
of the approximate Dirichlet-Neumann operators defined in \emph{Section 2} and above,
with $b-$periodic depth profiles  
with vanishing depth at integer multiples of $z = b$. 
Also, the domains we consider in the next section are symmetric in $z$ so that
the eigenfunctions of the approximate Dirichlet-Neumann operators are either even or odd, satisfying
Neumann and Dirichlet boundary conditions respectively at $z = 0$, $b$.

\section{Transverse and longitudinal modes in triangular cross-sections}\label{Sec3p3}
\label{sec:3}
Transverse and longitudinal modes can be calculated explicitly only
for special geometries of the channel cross-sections.
In this section we compare some exact results for triangular channels by
Lamb \cite{lamb1932hydrodynamics} \textit{Art. 261},  Macdonald \cite{macdonald1893waves},
Greenhill \cite{greenhill1887wave}, Packham \cite{packham1980small}, and Groves \cite{groves1994hamiltonian},
to results obtained using the approximate Dirichlet-Neumann operators defined in the previous sections.

\subsection{Transverse modes for triangular cross-section: right isosceles triangle}
\label{sec:3.1}
\begin{figure}
    \includegraphics[scale=0.45]{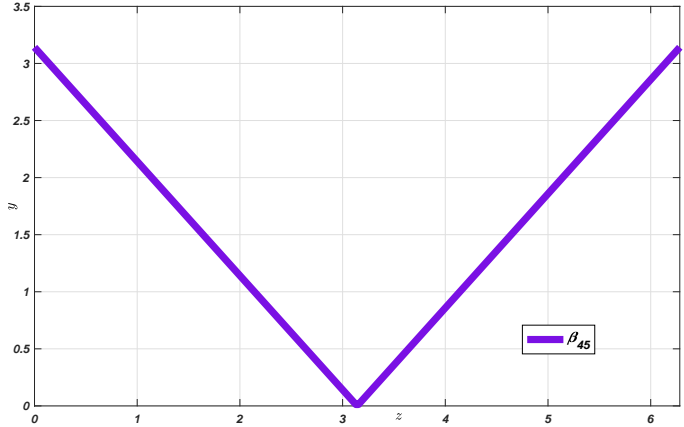}
    \caption{Right isosceles triangular cross-section of a straight channel  with the right angle at the bottom, see equation \protect\eqref{beta45}} \label{Figbeta45} 
\end{figure}

The first geometry we consider corresponds
to a uniform straight channel with right isosceles triangle as a cross-section, see \figurename{ \ref{Figbeta45}}.
The cross-section $\Omega= \Omega_{B}$ is as in  \eqref{omega-bounded} and the bottom is
at $y = \beta_{45}(z)$. The minimum and maximum heights of the the fluid domain are
$h_m = 0$ and $h_{M}=\pi$ respectively
The  channel width is  $b=2\pi$, and
\begin{equation} \label{beta45}
\beta_{45}(z) =\left\lbrace \begin{array}{l}
-z + \pi \text{ } \text{ in }  0 \leq z < \pi \\
z -\pi \text{ } \text{ in }  \pi \leq z \leq 2\pi. \\
\end{array} \right.
\end{equation}

\begin{figure}      
        \includegraphics[scale=0.19]{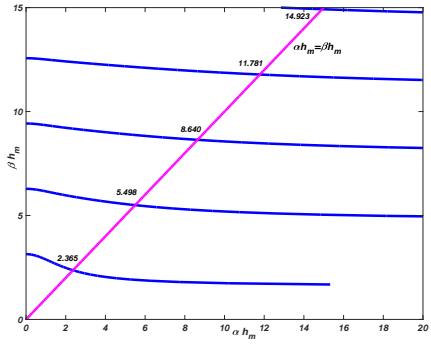}
           \caption{Even modes.  Intersection of curves \protect\eqref{kzero246} and \protect\eqref{curva246}}            
        \label{fig:Even45}
    \end{figure}

Normal modes for this channel were obtained by Kirchhoff, see Lamb \cite{lamb1932hydrodynamics},
\textit{Art. 261}, and include symmetric and antisymmetric modes.
The  symmetric transverse modes, see \eqref{TransverseMode}, are given by potentials
$\phi(x,y,z,t) = \psi(y,z)\cos(\omega t)$ with 
\begin{equation}\label{SymModes}  
\psi(y,z) = A[\cosh(\alpha z)\cos(\beta y) + \cos(\beta z)\cosh(\alpha y)].
\end{equation}
It can be checked that
$\frac{\partial \phi}{\partial y} \mid_{y=h_M}$ and
$\phi$ are symmetric with respect to the $z = 0$ axis. Also, $\phi$
in harmonic in the quarter plane $y \geq |z|$, and
satisfies the rigid wall boundary condition
\begin{equation}
\frac{\partial \phi}{\partial \hat{n}}=0  \quad \text{at} \quad y= |z|.
\end{equation}
To impose the boundary condition at the free surface $y = h_M$, we
use the first two equations of \eqref{Ebc} to obtain
\begin{equation}\label{SymModesSurface}
\omega^2  \psi = g \frac{\partial \psi}{\partial y}   \quad \text{at} \quad y= h_M.
\end{equation}\label{kzero246}
Combining with \eqref{SymModes} we have the conditions
\begin{equation}
  \label{kzero246}
\alpha^2 - \beta^2 =0,
\end{equation}
and
\begin{equation}\label{omega246}
\omega^2 \cosh(\alpha h_M)= g\alpha \sinh(\alpha h_M),
\quad
\omega^2\cos(\beta h_M)=-g\beta \sin(\beta h_M)=0,
\end{equation}
or equivalently
\begin{equation}\label{curva246}
\alpha h_M \tanh (\alpha h_M) + \beta h_M \tan (\beta h_M)=0.
\end{equation}
The values of $\alpha$, $\beta$ are determined by the intersections of the curves
\eqref{kzero246} and \eqref{curva246}, see
\figurename{ \ref{fig:Even45}}. There is an infinite number of solutions,
$h_M \alpha_j$, $j= 0,2,4,\ldots$,  with $\alpha_{j}< \alpha_{j'}$ if $j<j'$. The
corresponding frequencies $\omega_j$ are obtained by \eqref{omega246}.
The first values of $h_{M}\alpha_i, \omega_i$ are shown in Table 1. 

\begin{figure}       
        \includegraphics[scale=0.2]{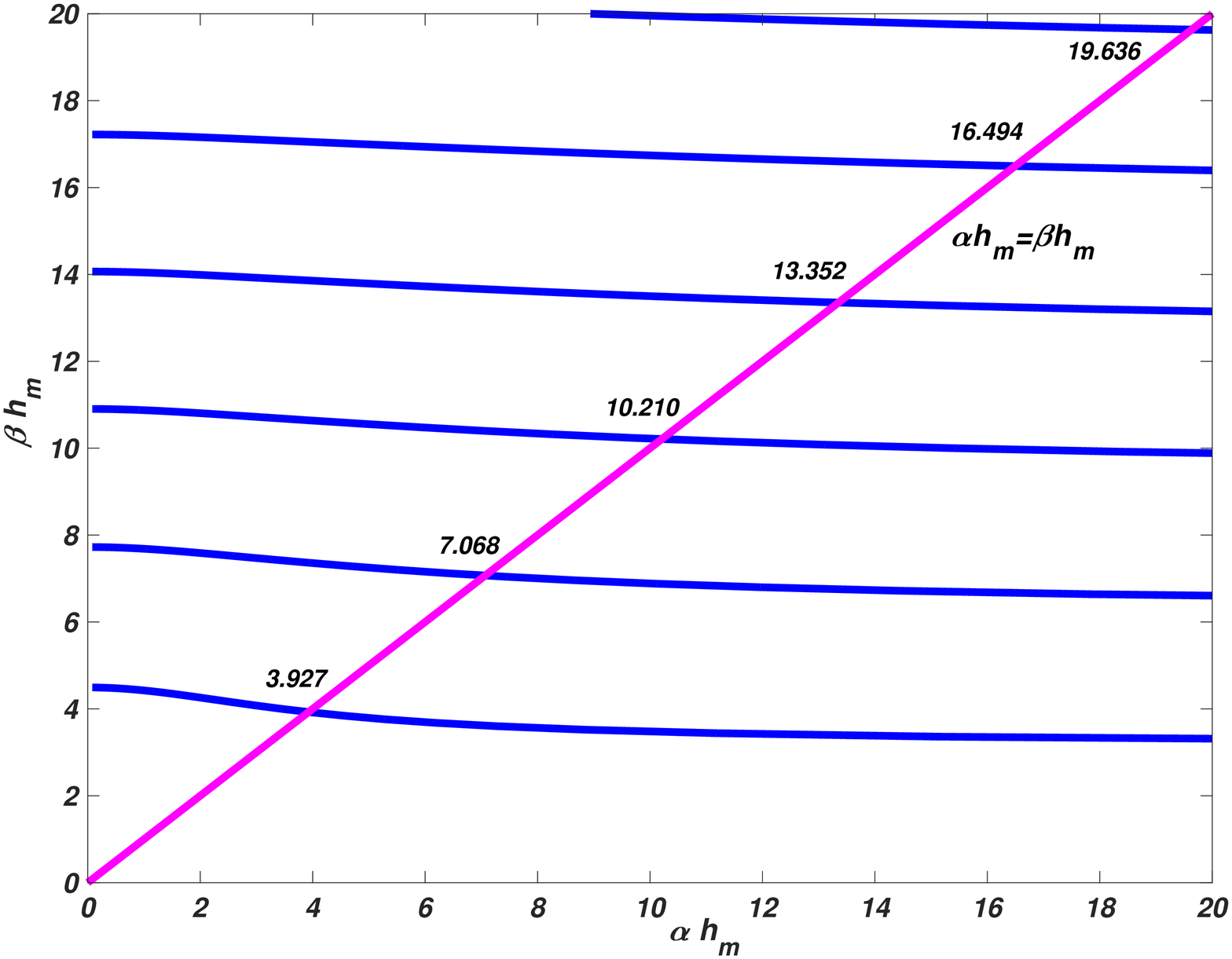}
           \caption{Odd modes.  Intersection of curves \protect\eqref{kzero357} and \protect\eqref{curva357}}
                \label{fig:Odd45}
    \end{figure}

To obtain
the antisymmetric modes we use the potentials
$\phi(x,y,z,t) = \psi(y,z)\cos(\omega t)$ with
\begin{equation}\label{AsymModes}
\psi(y,z)= B[\sinh(\alpha z) \sin(\beta y)+ \sin(\beta z) \sinh(\alpha y)].
\end{equation}
We check that $\phi $
satisfies the rigid wall
boundary conditions at $y = |z|$ and is harmonic in the quadrant $y \geq |z|$.
We also check that $\frac{\partial \phi}{\partial y} \mid_{y=h_M}$
is antisymmetric with respect to the $z = 0$ axis.
Imposing the free surface boundary conditions \eqref{Ebc} to \eqref{TransverseMode} we obtain
\eqref{SymModesSurface}. Then \eqref{AsymModes}
leads to the conditions
\begin{equation}\label{kzero357}
\alpha^2 - \beta^2=0
\end{equation}
and
\begin{equation}\label{omega357}
\omega^2 \sinh(\alpha h_M)= g\alpha \cosh(\alpha h_M),
\quad \omega^2 \sin(\beta h_M)= g\beta \cos(\beta h_M),
\end{equation}
or
\begin{equation}\label{curva357}
\alpha h \coth(\alpha h_M)= \beta h_M \cot(\beta h_M).
\end{equation}
The values of $\alpha$, $\beta$ are determined by the intersections of
the curves \eqref{kzero357} and \eqref{curva357}, see \figurename{ \ref{fig:Odd45}}. 
There is an infinite number of solutions
$h_M\alpha_{j}$, $j={1,3,5...}$ with $\alpha_{j}< \alpha_{j'}$ if $j<j'$.
The corresponding frequencies $\omega_{j}$ are given by \eqref{omega357}, see Table 1.

By the first two equations of \eqref{Ebc}, and the form of $\phi$,
the free surface corresponding to the above symmetric and antisymmetric
satisfies
\begin{equation}\label{freeS}
  \eta(x,z,t) = - \frac{1}{g} \omega \psi(y,z)\mid _{y=h_M} \sin \omega t =
  \frac{1}{\omega}  \frac{\partial \psi (y,z)}{\partial y} \mid _{y=h_M} \sin \omega t.
\end{equation}
The amplitude of $\eta $ is therefore $\omega^{-1}  \frac{\partial \psi (y,z)}{\partial y} \mid _{y=h_M}$.

\begin{table}
    \caption{Frequencies of modes for channel of \figurename{ \ref{Figbeta45}} these values are associated to graphic roots in \figurename{ \ref{fig:Even45}} and \ref{fig:Odd45} We use $h_M=\pi$}
    \label{tab:table3p1}       
    \begin{tabular}{cccccccc}
        \noalign{\smallskip}\hline\noalign{\smallskip}   
         &$i=0$ & $i=1$& $i=2$ & $i=3$ & $i=4$& $i=5$ & $\cdots$ \\
        $\alpha_i h_M$ & 2.365 & 3.927 & 5.498 & 10.210 & 11.781& 16.494 & $\cdots$\\
        $\omega_i$ &4.8624   & 7.8261  &4.1413 & 5.6445  &6.0622 & 7.1684  & $\cdots$\\
        \noalign{\smallskip}\hline
    \end{tabular}
\end{table}

\begin{figure}
    \includegraphics[scale=0.37]{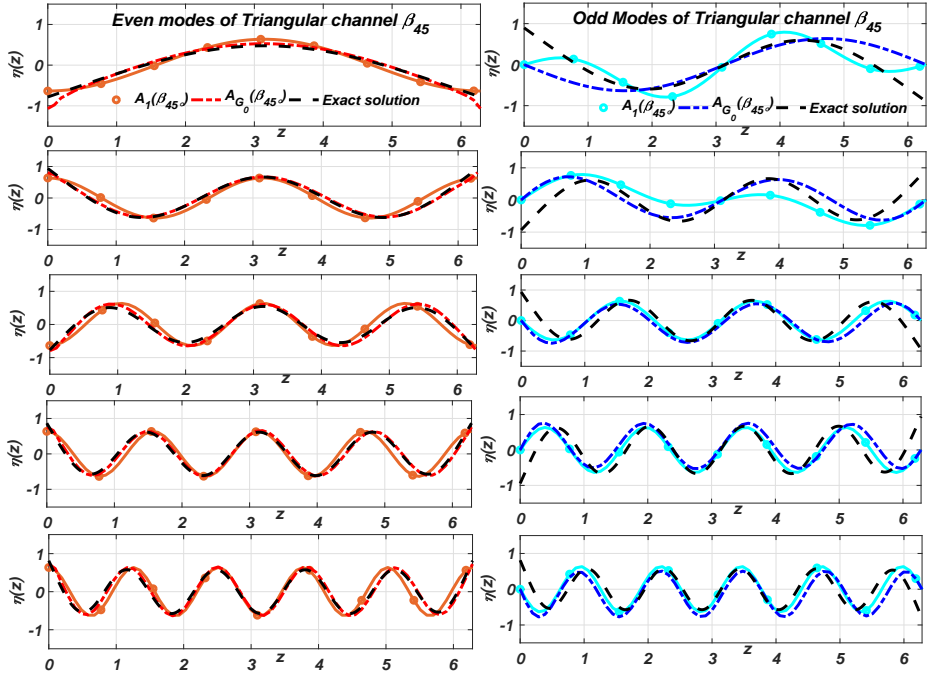}
    \caption{Transverse modes of a channel with isosceles triangular cross-section illustrated in  \figurename{ \protect\ref{Figbeta45}}. Left: Even modes. Orange (circle-line): operator $\mathcal{A}_{1}(\beta_{45^{\circ}})$.Red (dot-dashed-line): operator  $\mathcal{A}_{G_0}(\beta_{45^{\circ}})$. Black dashed lines: exact solutions  given by \protect\eqref{SymModes}  and the values in Table 1. Right: Odd modes. Cian (circle-line): operator $\mathcal{A}_{1}(\beta_{45^{\circ}})$.  Blue (dot-dashed-line): operator  $\mathcal{A}_{G_0}(\beta_{45^{\circ}})$. exact solutions  given by \protect\eqref{AsymModes}  and the values in Table 1}
    \label{T45evenandodd}
\end{figure}

In \figurename{ \ref{T45evenandodd}}
we compare the surface amplitude of
the exact symmetric and antisymmetric modes found above to the
surface amplitudes obtained by computing numerically
the eigenfunctions of the approximate Dirichlet-Neumann operators
$ {\mathcal{A}}_{G_{0}}(\beta_{45^{\circ}})$ of \eqref{AG0},
and $\mathcal{A}_1(\beta_{45^{\circ}})$ 
with $2 \pi-$periodic boundary conditions. The surface amplitude $\eta$ is obtained
by $\eta = \omega^{-1} \mathcal{A}(\beta)f$, with
$ {\mathcal{A}}(\beta)$ representing each of the two approximate Dirichlet-Neumann operators.
This expression is analogous to \eqref{freeS}.

\figurename{ \ref{T45evenandodd}} suggests
good quantitative agreement between the exact even modes and the even modes
obtained by $\mathcal{A}_{G_0}$, $\mathcal{A}_1(\beta)$,
with some discrepancy near the boundary for the first mode.
The $\mathcal{A}_{G_0}$ modes seem closer to the exact ones in the interior.
For the odd modes, the two approximate operators lead to vanishing amplitude
at the boundary. This is a consequence of
the parity considerations of the previous section. On the other hand,
exact odd modes have non-vanishing values at the boundary, in fact they appear to have
local extrema at the boundary.
This leads to a discrepancy between exact and approximate
modes at the boundary. The first two modes of
$\mathcal{A}_{G_0}$, $\mathcal{A}_1(\beta)$ and of the exact approach differ quite significantly
also in the interior
of the domain, with the $\mathcal{A}_{G_0}$ modes being somewhat closer to the exact ones. 
For higher modes, the discrepancy at the boundary persists, but the values of the interior are close
for all four sets of modes.


The procedure for obtaining
the exact modes does not require any conditions on the value
of the potential at the intersection of the free boundary and the rigid wall.
Also, the free surface is described by a value of the $y-$coordinate, and this allows
us to define $\eta(z)$ for all real $z$, in particular
we determine the fluid domain by computing the intersection
of the graph of $\eta$ with the rigid wall (\figurename{ \ref{T45evenandodd}} only shows $\eta(z)$, $z \in [0,2\pi]$).
The exact approach leads then to a more realistic motion of the surface at the sloping beach.
This does not imply that the exact solutions are physical either, since the
boundary conditions at the free surface are not exact.
In contrast,
the odd modes of the approximate operators correspond to
pinned boundary conditions that are not expected to be physical.

\begin{figure}
    \includegraphics[scale=0.32]{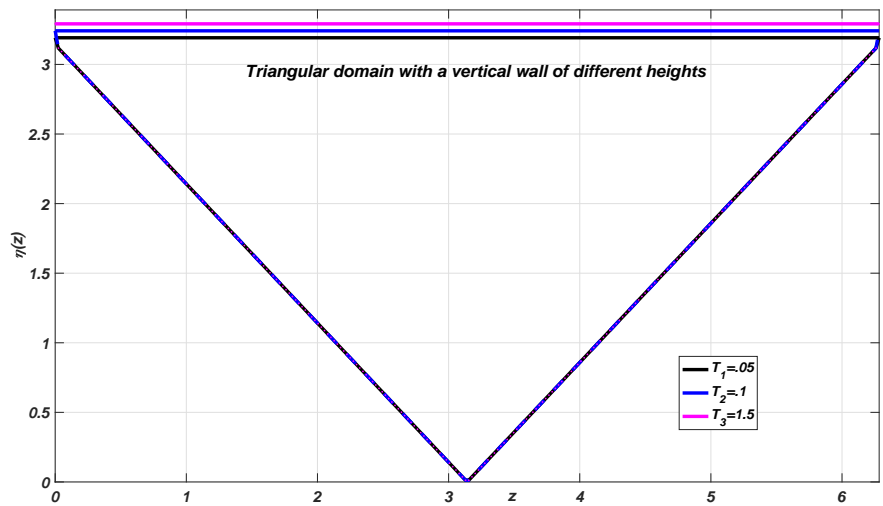}
    \caption{Cross section of the straight channels with
        lateral boundary $\Gamma_L$. Vertical walls of heights $T_1=0.05$ (black), $ T_2=0.1$ (red), $T_3=.15$ (blue)} \label{CSt1t2t3}
\end{figure}

\begin{figure}
   \includegraphics[scale=0.33]{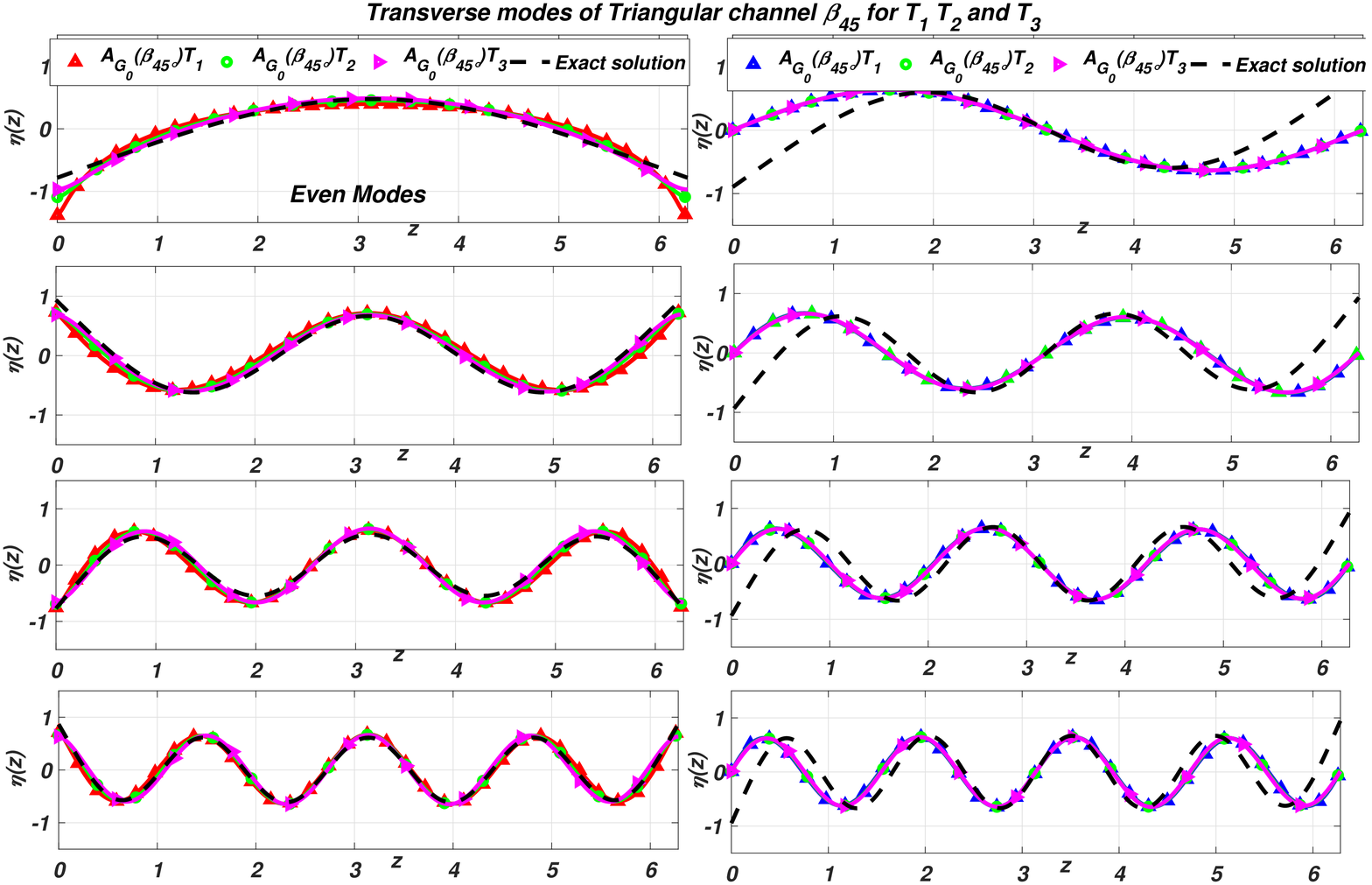}
    \caption{Transverse modes with $\mathcal{A}_{G_{0}}(\beta_{45})$  for the three cross sections of \figurename{ \protect\ref{CSt1t2t3}}, vertical segments of
        heights $T_1=.006$ (red), $ T_2=.15$ (green) , $T_3=.6$ (magenta)}
    \label{ModesT45t1t2t3OPD}
\end{figure}

We have also used the operators
$\mathcal{A}_{G_0}(\beta_{30^{\circ}})$ and $\mathcal{A}_{1}(\beta_{30^{\circ}})$
to compute the $2 \pi-$periodic
normal modes of domains obtained from the triangular channel
by adding an interval of extra depth, see \figurename{ \ref{CSt1t2t3}}.
We denote the added depth by $T$.
\figurename{ \ref{ModesT45t1t2t3OPD}}
indicate the convergence as $T $ vanishes of the modes obtained
for $\mathcal{A}_{G_0}(\beta_{45})$
for $T > 0 $ to the
corresponding modes obtained for the triangular domain $ T = 0$.
Similar results were obtained for the other operators.
Also,
even modes satisfy a Neumann boundary condition at $z = 0$, $2 \pi$,
and the interval of length
$T$ can be interpreted also as the height of a vertical wall at $z= 0$, $2 \pi$.

Convergence to the triangular domain modes indicates more clearly that normal modes of
the approximate, $2 \pi-$periodic
operators used for a triangular domain are limiting cases of operators defined
for a $2 \pi-$periodic depth profile, with depth that does not vanish anywhere.
In that case the definition and computations of the Dirichlet-Neumann operator
follow the construction of \cite{CGNS}, but can not take into account the sloping
beach boundary.
Note also that
the $2 \pi-$periodic modes obtained using the $2 \pi-$periodic
$\mathcal{A}_{G_0}(\beta_{30^{\circ}})$, $\mathcal{A}_{1}(\beta_{30^{\circ}})$, are special cases of the Floquet-Bloch modes for
the $2 \pi-$periodic depth profile see \textit{arXiv:1706.07417} submitted manuscript by W. Craig, M. Gazeau, C. Lacave, C. Sulem.
for a study of the
the band structure and  Floquet-Bloch modes of $\mathcal{A}_{1}(\beta)$ with another
periodic profile.

%

\subsection{Longitudinal modes for triangular cross-sections: isosceles triangle with unequal angle of $120^\circ$ }
\label{sec:3.2}
\begin{figure}
    \includegraphics[scale=0.32]{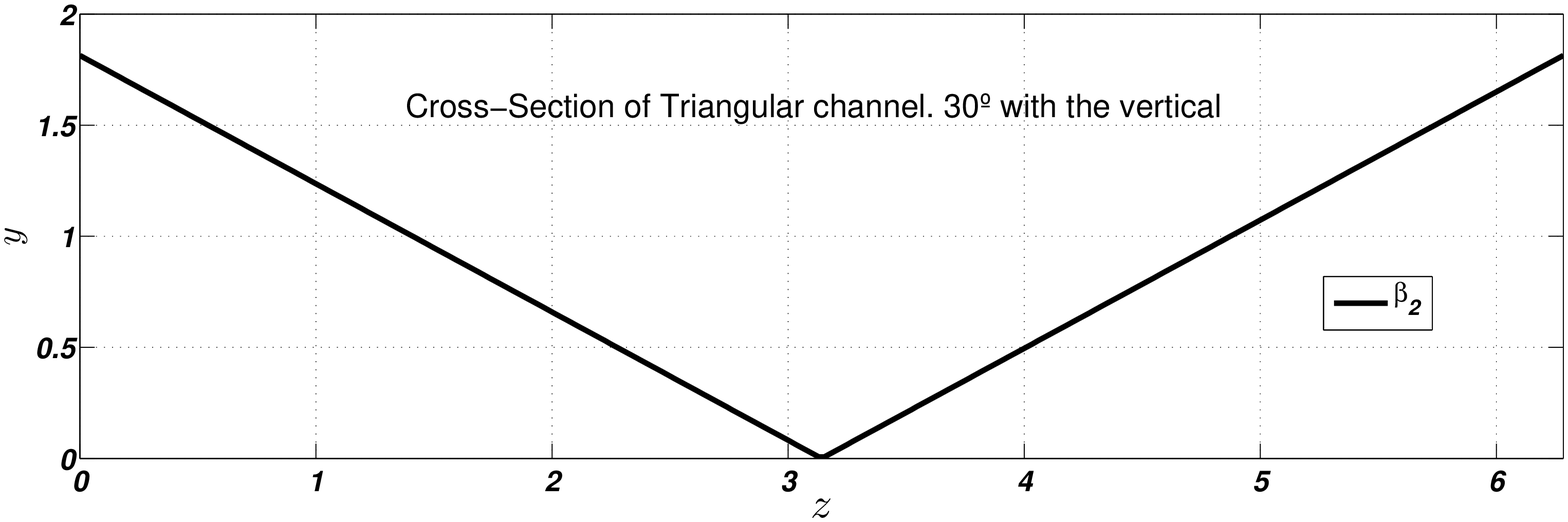}
    \caption{Straight channel with  isosceles triangular cross-section with the unequal angle at the bottom of $120^{\circ}$ , see equation \protect\eqref{beta30}}  
    \label{Figbeta30}
\end{figure}

A second geometry with exact longitudinal modes was
considered by Macdonald \cite{macdonald1893waves}, Packham, \cite{packham1980small}, see also
Lamb \cite{lamb1932hydrodynamics}, \textit{Art. 261}. This geometry
corresponds to a uniform straight channel with  isosceles triangle cross-section
with a  unequal angle at the bottom of $120^{\circ}$, as illustrated in \figurename{ \ref{Figbeta30}}. In this case we will examine longitudinal modes.

We consider the cross-section $\Omega= \Omega_{B}$, as in \eqref{omega-bounded} and
the bottom  $\beta_{30^{\circ}}(z)$ of \eqref{beta30}.
The  channel width is  $b=2\pi$ and the maximum and minimum heights of the fluid domain are
$h_{M}=\frac{\pi}{\sqrt{3}}$ and $h_m = 0$ respectively.
The cross-section profile is given by
\begin{equation} \label{beta30}
\beta_{30^{\circ}}(z) =\left\lbrace \begin{array}{l}
\frac{-1}{\sqrt{3}}z + \frac{\pi}{\sqrt{3}},   0 \leq z < \pi \\
\frac{1}{\sqrt{3}}z -\frac{\pi}{\sqrt{3}},  \pi \leq z \leq 2\pi \\
\end{array} \right. , z \in [0, 2\pi].
\end{equation}

The exact  solutions for symmetric modes, see Packham \cite{packham1980small},
and Groves \cite{groves1994hamiltonian} are as follows.

The $0-$mode is described by a velocity potential $\phi$ of the form \eqref{LongMode} with
\begin{eqnarray}\label{lowestmode}
\psi(z,y) &=& A[\cosh (\kappa(y-h_M))+
\frac{\omega^2 \kappa^2}{g\kappa}\sinh(\kappa(y-h_{M}))\\ \nonumber
& &+ 2\cosh(\frac{\sqrt{3}\kappa (z-\pi)}{2})\\ \nonumber
& &\times \lbrace \cosh(\kappa (\frac{y}{2}+ h_M)) -
\frac{\omega^2 \kappa^2}{g\kappa}\sinh(\kappa (\frac{y}{2}+ h_M))\rbrace ],
\end{eqnarray}
and
\begin{equation}
\label{lowestmodefreq}
\omega^2= \frac{3g}{4\kappa}\coth(\frac{3\kappa h_M}{2})\lbrace1
+ (1-\frac{8}{9} \tanh^2(\frac{3\kappa h_M}{2}))^{\frac{1}{2}}\rbrace.
\end{equation}

The remaining symmetric modes $2,4,6,8,...$ are
described by a velocity potential $\phi$ of the form \eqref{LongMode} with
\begin{eqnarray}\label{sym30}
\psi(y,z) &=& A[\lbrace \cosh(\alpha(y-h_M)) +\frac{\omega^2 \kappa^2}{g\kappa} \sinh(\alpha(y-h_M)) \rbrace \cos(\beta(z-\pi)) \\ \nonumber
& &+ 2\cosh(\frac{\sqrt{3}\alpha (z-\pi)}{2}) \cos(\frac{\sqrt{3}\beta y}{2}) \cos(\frac{\beta (z-\pi)}{2})\\\nonumber
& &\times\lbrace \cosh(\alpha(\frac{y}{2}+h_M)) - \frac{\omega^2 \kappa^2}{g\kappa} \sinh(\alpha (\frac{y}{2} + h_M)) \rbrace \\\nonumber
& & -2 \sinh(\frac{\sqrt{3}\alpha (z-\pi)}{2}) \sin(\frac{\sqrt{3}\beta y}{2} \sin(\frac{\beta (z-\pi)}{2}))\\\nonumber
& & \times \lbrace \sinh(\alpha(\frac{y}{2}+h_M)) - \frac{\omega^2 \kappa^2}{g\kappa}\cosh(\alpha(\frac{y}{2}+h_M))   \rbrace],
\end{eqnarray}
and
\begin{equation}\label{omegasym30}
\omega^2 =\frac{g\alpha}{\kappa^2}\left[
\frac{\frac{\beta}{\alpha}\sqrt{3}(\cosh(3\alpha h_M)- \cos(\sqrt{3}\beta h_M))}{\frac{\beta}{\alpha}\sqrt{3}\sinh(3\alpha h_M)-3 \sin(\sqrt{3}\beta h_M)}
\right].
\end{equation}
The above potentials are harmonic and satisfy the rigid wall boundary conditions.

\begin{figure}
    \includegraphics[scale=0.37]{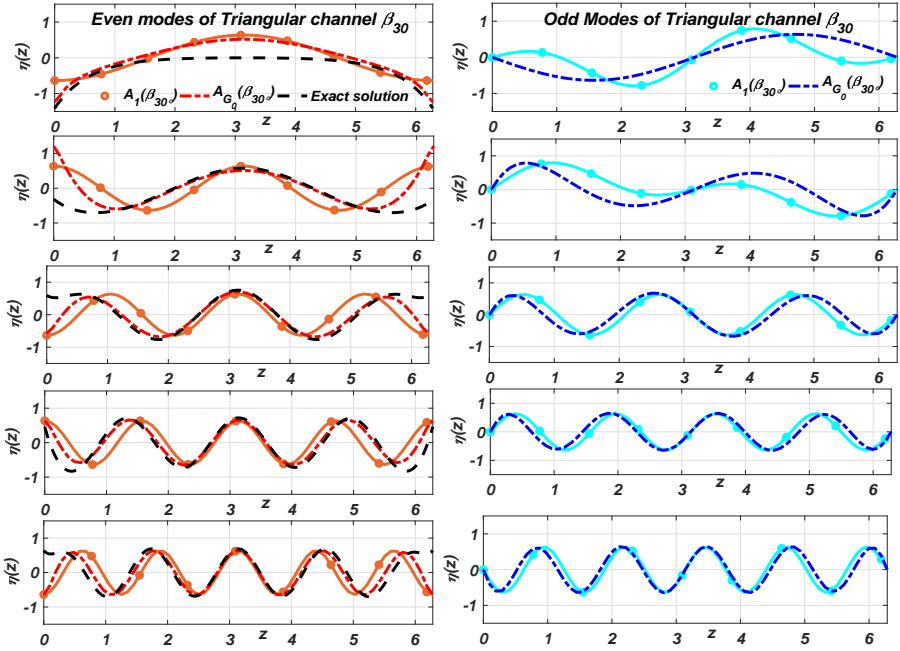}
    \caption{Longitudinal modes of a channel with isosceles triangular cross-section illustrated in \figurename{ \protect\ref{Figtrian30}}. Left: Even modes  with $\kappa=2$. Orange(circle-line): operator $\mathcal{A}_{1}(\beta_{30^{\circ}})$.
          Red (dot-dashed-line): operator  $\mathcal{A}_{G_0}(\beta_{30^{\circ}})$.
          Black dashed lines: exact solutions  given by \protect\eqref{sym30} and \protect\eqref{omegasym30} and the values in Table 2.
          Right: Odd modes  with $\kappa=2$. Cian (circle-line): operator $\mathcal{A}_{1}(\beta_{30^{\circ}})$ with $\kappa=2$.
          Blue (dot-dashed-line): operator  $\mathcal{A}_{G_0}(\beta_{30^{\circ}})$. }
    \label{30evenoddModes}
\end{figure}

The first two equations of motion of \eqref{Ebc}, \eqref{LongMode}, and \eqref{sym30} lead to
\begin{equation}\label{alfabetakappa}
\alpha^2- \beta^2= \kappa^2,
\end{equation}
and
\begin{eqnarray}\label{relalfabetakappa}
& &\left( \frac{\beta}{\alpha}\right) ^2 \cosh(3\alpha h_M)\cos(\sqrt{3}\beta h_M)\\ \nonumber
&-&\frac{1}{4}\left( \frac{\beta}{\alpha}\right)  \sqrt{3} \left\lbrace  1-\left( \frac{\beta}{\alpha}\right) ^2 \right\rbrace \sinh(3\alpha h_M)\sin(\sqrt{3}\beta h_M)\\\nonumber
& &-\frac{1}{4}\left[\left\lbrace  3 + 5\left( \frac{\beta}{\alpha}\right) ^2 \right\rbrace  - \left\lbrace  3 +\left(  \frac{\beta}{\alpha} \right) ^2\right\rbrace  \cos^2(\sqrt{3}\beta h_M)\right] =0.
\end{eqnarray}
In \figurename{ \ref{30evenoddModes}} we show the symmetric longitudinal modes derived
with the values of $\alpha$ and $\beta$ obtained
from relations \eqref{alfabetakappa} and \eqref{relalfabetakappa}
for $\kappa=2$, see also Table 2.

By the first two equations of \eqref{Ebc} and \eqref{LongMode},
the free surface corresponding to the above
modes can be computed by
\begin{equation}
  \label{freeS-kappa}
  \eta(z,t)=
-  \frac{1}{\omega}  \frac{\partial \psi (y,z)}{\partial y} \mid _{y=h_M} \sin \omega t.
\end{equation}
The amplitude of $\eta $ is therefore $\omega^{-1}  \frac{\partial \psi (y,z)}{\partial y} \mid _{y=h_M}$.

\begin{figure}
    \includegraphics[scale=0.35]{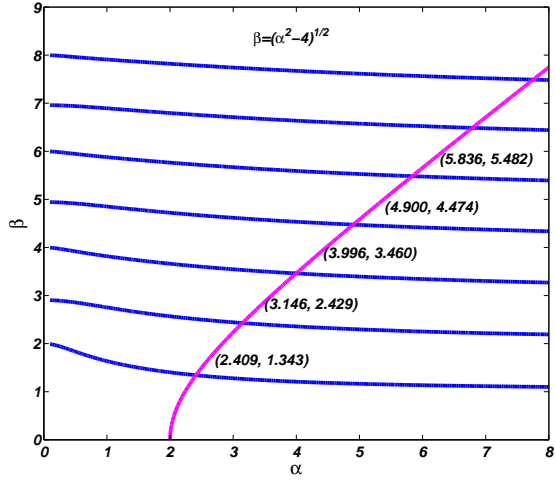}
    \caption{Graphs for the determination of $\alpha$ and $\beta$ for the
        modes $2$, $4$, $6$, $8$, and $10$ using $\kappa=2$,
        $h_M=\frac{\pi}{\sqrt{3}}$} \label{30locus}
\end{figure}

\begin{table}
    \caption{Frequencies of modes of channel of \figurename{ \ref{Figbeta30}} these values are associated to graphic roots in \figurename{ \ref{30locus}}. We use $h_M=\frac{\pi}{\sqrt{3}}$}
    \label{tab:table3p1}       
    \begin{tabular}{ccccccc}
        \hline\noalign{\smallskip}
      &$i=2$ & $i=4$& $i=6$ & $i=8$ & $i=10$ & $\cdots$ \\
        \noalign{\smallskip}\hline\noalign{\smallskip}   
    $\alpha_i$ & 2.409 & 3.146 & 3.996 & 4.900 & 5.836 & $\cdots$\\
    $\beta_i$ & 1.343 & 2.429 & 3.460 & 4.474 & 5.482 & $\cdots$\\
    $\omega_i$ &2.4297   & 2.7764  &3.1289 & 3.4651  &3.7813  & $\cdots$\\
        \noalign{\smallskip}\hline
    \end{tabular}
\end{table}

In \figurename{ \ref{30evenoddModes}}
we compare the surface amplitude of
the exact symmetric modes  to the
surface amplitudes obtained by computing numerically the eigenfunctions of
the approximate Dirichlet-Neumann operators
$ \mathcal{A}_{G_{0,\kappa}}(\beta)$, $ \mathcal{A}_{1,\kappa}(\beta)$ of
\eqref{Aproxgorro-kappa}, \eqref{A1en3D-kappa} respectively,
with
$\beta$ as in \eqref{beta30}.
We use  $\kappa=2$. To compute the
eigenfunctions of ${\mathcal{A}}_{G_{0,\kappa}}(\beta)$,  $ \mathcal{A}_{1,\kappa}(\beta)$
numerically we
use $2 \pi-$periodic boundary conditions.
Also, given a computed eigenfunction $f$ of  ${\mathcal {A}} = {\mathcal{A}}_{G_{0,\kappa}}(\beta)$
or  $ \mathcal{A}_{1,\kappa}(\beta)$ the surface amplitude $\eta$
is given by $\eta = \omega^{-1}  \mathcal{A}$.  This is analogous to \eqref{freeS-kappa}.

By \figurename{ \ref{30evenoddModes}} we see some discrepancies
between the first exact even mode and the first even
${\mathcal{A}}_{G_{0,\kappa}}(\beta)$ mode. For higher even modes,
the ${\mathcal{A}}_{G_{0,\kappa}}(\beta)$ modes are close to the exact
modes in the interior, but show discrepancies at the boundary.
Also the even modes obtained by ${\mathcal{A}}_{G_{0,\kappa}}(\beta)$,  $ \mathcal{A}_{1,\kappa}(\beta)$
are generally close both in the interior and the boundary, with more pronounced
discrepancies for the first and second modes.

To our knowledge there are no exact solutions reported in the literature for odd modes.
Odd modes obtained
with the approximate Dirichlet-Neumann operators
$\mathcal{A}_{G_{0,\kappa}}(\beta_{30^{\circ}})$, $\mathcal{A}_{1, \kappa}(\beta_{30^{\circ}})$
are shown in \figurename{ \ref{30evenoddModes}}.


\section{Discussion}
\label{sec:4}

We have studied linear water wave modes in channels with
variable depth, choosing depth geometries and models with known exact results.
The main goal was to test simplifications of the lowest order variable depth
Dirichlet-Neumann operator for variable
depth. The exact results involve slopping beach geometries,  
while the approximate Dirichlet-Neumann operators we use
are seen to be limits of approximate Dirichlet-Neumann operators
for periodic topographies with nowhere vanishing depth.
This observation suggests that the problems we compare are not equivalent.
Despite this fact we see reasonable agreement in the interior of the domain,
with most discrepancies at the boundary of the
free surface.

In the case of 2-D even modes the approximate
operators yield good approximations of the exact modes even at the boundary.
In the case of 2-D odd modes, the 
approximate Dirichlet-Neumann operators impose Dirichlet boundary
conditions and miss the boundary behavior of the exact modes.
In general, the exact modes seem to have local extrema at the boundary, and this
may explain why the even modes of the periodic approximate Dirichlet-Neumann operators may
give a better fit. We also think that Neumann
boundary conditions, allowing odd modes, may give better approximations.
In the case of 3-D longitudinal waves, the exact approach only yields even modes.
The approximate operators give reasonable approximations of the exact modes
in the interior
of the domain, but can miss the boundary behavior. We suspect that
Neumann boundary conditions may also be more appropriate.


\begin{acknowledgements}
We would like to thank especially Professor
Noel Smyth for many helpful comments.
R. M. Vargas-Maga\~na was supported by Conacyt Ph.D. scholarship 213696.
P. Panayotaros and R. M Vargas-Maga\~na also acknowledge partial support from grants SEP-Conacyt 177246 and PAPIIT IN103916. This material is based upon work supported by the National Science Foundation under Grant No. DMS-1440140 while R. M. Vargas-Maga\~na was in residence at the Mathematical Sciences Research Institute in Berkeley, California, during the Fall 2018 semester.

\end{acknowledgements}

\section*{Appendix A}

We present some computations related to 
symmetrization, parity, and
the operator $D \tanh (h(\cdot) D)$.

The notion of adjoint $\mathcal{A}^*$
applies to operators $\mathcal{A}:D(\mathcal{A}) \subset L^2 \rightarrow  L^2 $,
with $D(\mathcal{A})$ dense in $L^2 = L^2(\mathbb{R};\mathbb{R})$.
Operators that map real-valued functions to real-valued (resp. imaginary-valued) functions
will be denoted as {\it real} (resp. {\it imaginary}) operators.
Imaginary operators map $D(\mathcal{A}) \subset L^2$ to $ i L^2 $.
The adjoint and symmetrization of a real operator is real.
We extend the definition of the adjoint to imaginary operators
linearity by requiring $\langle \mathcal{A} f, g \rangle = \langle f, \mathcal{A}^* g \rangle $,
for all $f \in D(\mathcal{A})$. Letting $\mathcal{B} = i \mathcal{A}$,
we have $\mathcal{A}^* = -i \mathcal{B}^*$.
We note that $D$ and $\tanh (h(x) D) f$ are imaginary, and therefore
$ D\tanh (h(x) D) f$ is real. Similarly, we check that operators
$\mathcal{A}_1$, $\mathcal{A}_2$,
$\mathcal{A}_{1,\kappa}(\beta)$, $\mathcal{A}_{G_{0,\kappa}}(\beta)$ are also real.

Also, for $\beta$, $h$ even we check that
$\mathcal{A}_1$, $\mathcal{A}_2$, $D\tanh (h(x) D)$,
$\mathcal{A}_{1,\kappa}(\beta)$, and $\mathcal{A}_{G_{0,\kappa}}(\beta)$
preserve parity, i.e.
map even (resp. odd) real-valued functions to even (resp. odd) real-valued functions.
This follows by examining the various operators appearing in the respective definitions
and their compositions.

For instance, the operator $D $ maps even (resp. odd)
real-valued functions to odd (resp. even)
imaginary-valued functions. Also, by the definition of $\tanh(h(x)D)$ on the line, 
\begin{eqnarray} \nonumber
  g_1(x) & = & \tanh(h(x)D) \cos k x =
  \frac{1}{2} [\tanh(h(x)k) e^{i k x} + \tanh(h(x) (- k)  e^{- ik x}] \\ \nonumber
  &  = & i \tanh(h(x)k)  \sin k x,
\end{eqnarray}
\begin{eqnarray} \nonumber
  g_2(x) & = & \tanh(h(x)D) \sin k  x =
  \frac{1}{2 i} [\tanh(h(x) k ) e^{i k x} - \tanh(h(x) (- k))  e^{-i k  x}] \\  \nonumber
  & = & - i \tanh(h(x) k)  \cos k x.
\end{eqnarray}
Then
$h $ even implies 
$g_1(-x) = - g_1(x)$, and $g_2(-x) = g_2(x)$, for all $x$. Therefore
$\tanh(h(x)D) $  maps even (resp. odd) real-valued functions to odd (resp. even)
imaginary-valued functions, and $D \tanh(h(x)D)$ is real and preserves parity  
Similar calculations apply to $b-$periodic functions, e.g. with $k$
integer if $b = 2 \pi$.
Operators $ \mathcal{A}_{1,\kappa}(\beta) $ of \eqref{A1en3D-kappa}
and $ \mathcal{A}_{G_{0,\kappa}}(\beta) $
of \eqref{Aproxgorro-kappa} are compositions of real operators that preserve parity.

We now consider the operator $\mathcal{A}_{G_0}$ up to order one in $\beta$. We
have
\begin{eqnarray}\label{dertanh}
  [D\tanh (h(x) D)f](x)
&=&-i\partial_x  (2 \pi)^{-1} \int_{\mathbb{R}} \tanh (h(x) k)\hat{f}(k) e^{ik x} d k \nonumber\\
  &=& (2 \pi)^{-1} [-i \int_{\mathbb{R}}[(\partial_x\tanh (h(x)k))\hat{f}(k) e^{i k x} d k \nonumber \\
      & & \qquad \qquad \qquad \qquad + \int_{\mathbb{R}} (\tanh (h(x) k))\hat{f}(k)k e^{i k x} d k] \nonumber\\
&=& (2 \pi)^{-1} [i\beta'(x) \int_{\mathbb{R}}  \text{sech}^2(h(x) k) k \hat{f}
      (k ) e^{i k x} d k \nonumber \\
      & & \qquad \qquad \qquad \qquad + \int_{\mathbb{R}} (\tanh (h(x) k))\hat{f}( k) k e^{i k x} d k] \nonumber\\
&=& i\beta'(x) [\text{sech}^2(h(x)D)Df](x) + [\tanh (h(x)D)Df](x),
\end{eqnarray}
using
\begin{eqnarray}
 \partial_x(\tanh (h(x) k)) &=&   \partial_x(\tanh(h(x) k)\partial_x (h(x) k)\nonumber \\
&=& \partial_x(\tanh(h(x) k))h'(x) k  \nonumber \\
&=&  -\text{sech}^2(h(x) k)\beta'(x) k, \nonumber
\end{eqnarray}
and $\partial_xh(x)=\partial_x(h_0-\beta(x))=-\beta'(x)$.
Furthermore
\begin{eqnarray}\label{PDOTaylorsech2}
[i\beta'(x)\text{sech}^2(h(x)D)Df](x)
&=& i\beta'(x) (2 \pi)^{-1} \int_{\mathbb{R}} [\text{sech}^2(h_0 k)+ O(\beta^2)] k \hat{f}(k)
e^{i k x} d k  \nonumber\\
&= &i\beta'(x)[\text{sech}^2(h_0D)D^2f](x) + O(\beta^2), \nonumber
\end{eqnarray}
and
\begin{eqnarray}\label{PDOTaylortanh}
  [\tanh (h(x)D)Df](x)
  &=&  \int_{\mathbb{R}}\left( \tanh(h_0 k)-\text{sech}^2(h_0 k)\beta(x) k+ O(\beta^2)\right) \hat{f}(k) k e^{i k x}
  \frac{d k}{2 \pi} \nonumber\\
&= &[\tanh(h_0D)Df](x) - \beta(x)[\text{sech}^2(h_0 D)D^2 f](x) + O(\beta^2). \nonumber
\end{eqnarray}
Therefore \eqref{dertanh} leads to
\begin{eqnarray}
  \label{D-tanhh(x)D}
  D \tanh (h(x)D) & =  &  \tanh (h_0 D)D + i\beta'\text{sech}^2(h_0 D)D - \beta \text{sech}^2(h_0 D)D^2 + O(\beta^2).
  \nonumber
  \end{eqnarray}
We also have
$ (\tanh (h_0 D))^* = - \tanh (h_0 D)$, $D^*= - D$, $(\text{sech}^2(h_0 D))^* = \text{sech}^2(h_0 D)$,
$ (i\beta \cdot)^* = i\beta \cdot$, $(\beta \cdot)^* = \beta \cdot$, so that
\begin{eqnarray}\label{sym(D-tanhh(x)D)}
  \textit{Sym}(D\tanh(h(x)D))  & =  & D\tanh (h_0 D)  +  \frac{1}{2}[ i\beta'\text{sech}^2(h_0 D)D
   - \beta \text{sech}^2(h_0 D)D^2  \nonumber \\  
    & & - i D \text{sech}^2(h_0 D) \beta' + D^2 \text{sech}^2(h_0 D) \beta] + O(\beta^2).                             
\end{eqnarray}
Operators $\text{sech}^2 (h_0 D)$,
$\beta \cdot$ (with $\beta $ even) are real and preserve parity,
while $i \beta' \cdot$, and $D$ are imaginary and reverse parity.
It follows that
the operator $\mathcal{A}_2$ of
\eqref{ExpansionAproxAG0Texto}
obtained by
truncating \eqref{sym(D-tanhh(x)D)} to $O(\beta^2)$ is real and preserves parity.

\end{document}